\documentclass[prl,twocolumn,superscriptaddress]{revtex4-1}

\usepackage{epsfig}
\usepackage{latexsym}
\usepackage{amsmath}
\usepackage{amssymb}
\usepackage{amsfonts}
\usepackage{graphicx}
\usepackage{wrapfig}
\usepackage[left]{lineno}
\usepackage{url}

\usepackage[normalem]{ulem}
\usepackage{color}

\definecolor{red}{rgb}{1,0,0}
\definecolor{green}{rgb}{0,1,0}
\definecolor{blue}{rgb}{0,0,1}

\begin{document}

\title{Pressure dynamics in the bottleneck flow of self-propelled particles}

\author{N. Colantuono}
\affiliation{Universidad de Buenos Aires, Facultad de Ciencias Exactas y Naturales, Departamento de Física. Buenos Aires, Argentina}%

\author{M. Ramdan Ferressini}
\affiliation{Universidad de Buenos Aires, Facultad de Ciencias Exactas y Naturales, Departamento de Física. Buenos Aires, Argentina}%

\author{I. Zuriguel}
%\email{iker@unav.es}
\affiliation{Departamento de F\'{\i}sica y Matem\'{a}tica Aplicada, Facultad de Ciencias,
Universidad de Navarra, Pamplona, Spain}

\author{D. R. Parisi}%
\affiliation{Instituto Tecnol\'ogico de Buenos Aires (ITBA), CONICET, Iguazú 341, 1437 C. A. de Buenos Aires, Argentina}

\author{G. A. Patterson}
\email{gpatters@itba.edu.ar}
\affiliation{Instituto Tecnol\'ogico de Buenos Aires (ITBA), CONICET, Iguazú 341, 1437 C. A. de Buenos Aires, Argentina}%

\date{\today}

\begin{abstract}
We present an experimental investigation of the pressure dynamics during the flow of self-propelled particles through narrow passages. When the ensemble is flowing, pressure fluctuates around a constant value that does not depend on the crowd size, suggesting that the orifice locally determines the dynamics in this scenario. On the contrary, when the system clogs, pressures are higher for larger crowd sizes, highlighting the importance of the whole collectivity in the process. Then, by correlating the pressure evolution with the exit time of the self-propelled particles, we discover that when a clog is resolved, pressure suddenly drops as a consequence of system reorganization. After this dramatic event, there is a sustained pressure growth over time that shows a square root dependence, compatible with the structural aging that has been proposed to be behind the broad tail distributions of clogging times. 
\end{abstract}

\maketitle

\section{Introduction}

When a system composed by discrete bodies flows through a bottleneck the flow may become interrupted due to the formation of clogs that arrest the flow \cite{SiloTO}. Then, the clog can be resolved by applying an external perturbation (if the particles are inert) \cite{Janda,hathcock2025stochastic} or by the intrinsic agents activity (if they are active) \cite{sheep}. The successive alternation of these two processes --clogging and unclogging-- leads to the characteristic intermittent flow observed in these scenarios, which range from the apparently simple silo discharge to pedestrian evacuation through a narrow gate \cite{zuriguel2014clogging}. In general, the flow in all these systems shows the same type of statistical features; i.e. an exponential distribution of the bursting times and a broad tail distribution --sometimes compatible with a power law-- of the clogs duration. 

The exponential distribution of the bursting times can be explained if the clog formation is a memoryless process with constant occurrence probability throughout all the duration of the burst \cite{zuriguel2005exponential,helbing2006exp,Roussel,masuda2014critical,Thomas}. However, the origin of the broad tail distribution of clogging times is not as well understood and two different theories have been proposed in the last years. The first one relied in the different energy necessary to resolve each of the configurations that are able to block a given geometrical configuration \cite{Nicolas}. The second attributed the broad tails to an aging process of the clogging configuration that would lead to a reduction of the unclogging probability with time \cite{Merrigan,GuerreroBulbul}. 

Interestingly, the only exceptions for the broad tail distribution in the unclogging times have been found in the flow of fish through an orifice \cite{fish} and pedestrian evacuations while keeping social distancing \cite{echeverria2022pedsocial}. In both scenarios, contacts among particles do not exist and therefore, there is not build up of pressure within the system. It is precisely the pressure —an ingredient that seems fundamental to understanding the clog destruction process— one of the variables whose behavior has been least studied in bottleneck flows. 

So far, we know that clogs last longer when increasing the driving force \cite{fasterisslower,Souzy_Marin_2022} or the crowd size \cite{patterson2017clogging}; both parameters in principle associated with the amount of pressure built up within the system. A similar behavior is found for the case of suspensions driven at constant pressures, as clogs last longer for higher pressure gradients. However, in the same scenario, if the liquid flow rate is imposed, the probability of breaking an arch is practically independent of the magnitude of it \cite{Souzy_Marin_2022,marin2024review}. Moreover, beyond some preliminary results for very competitive pedestrian flows \cite{pressurepeds}, and measurements in silos under flowing conditions \cite{expforce2}, there is not information of the pressure dynamics during intermittent flow. This is precisely the aim of this work. To this end, we use a system composed of self-propelled agents; a bidimensional scenario in which flow intermittency has already been reported \cite{patterson2017clogging,selfbottleneck}. Indeed, in \cite{patterson2017clogging} a transition from unclogged to clogged states when increasing the number of agents in the crowd, was well characterized. 

\section{Experimental setup}

\begin{figure}
\includegraphics{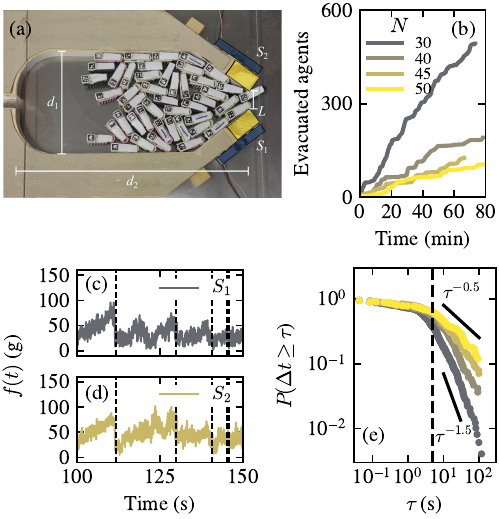}
\caption{\label{fig:experimental} (a) Experimental setup. Particles are introduced in the arena from the left, accumulate and exit through the small aperture of size $L = 30\ \mathrm{mm}$ at the right. This exit is made by two wooden sections (marked in yellow), on which the forces are measured by two force sensors, $S_1$ and $S_2$, marked in blue. The characteristic dimensions of the system are also displayed: $d_1 = 155\ \mathrm{mm}$, and $d_2 = 350\ \mathrm{mm}$. Each VDV is identified with a unique ArUco marker placed on its back. (b) Number of evacuated agents as a function of time for different number of total VDV within the arena as indicated in the legend. (c) Probability that $\Delta t$, the head time among consecutive VDVs, is longer than a given time $\tau$. The color code is the same as in (b). The vertical line shows the threshold that separates the bursting and clogging regimes. Two trend lines are included as references, corresponding to $P(\Delta t \geq \tau) \propto \tau^{-0.5}$ and $\propto \tau^{-1.2}$. (d) and (e) Temporal evolution of the force signals of the two sensors for a time window extracted from the experiment with $N_\mathrm{tot} = 30$. The vertical lines indicate the time at which a particle exits the arena.}
\end{figure}

We used commercially available vibration-driven vehicles (VDVs), known as Hexbug Nano, as an example of inertial self-propelled agent \cite{self3,self1,self4,self2}. Each vehicle measures $43\times15\times18\ \mathrm{mm}$, and its motion is generated by the vibration of an eccentric motor, rectified by a brush-like structure that acts as legs. The motor is powered by a 1.5 V battery. The direction of motion of these vehicles is influenced by collisions with other vehicles and can also be guided by the geometry of the container. The VDVs were placed inside an arena shaped similarly to a silo [see Fig.~\ref{fig:experimental}(a) for a snapshot including the characteristic dimensions]. The arena had two openings: one exit located at the end of a funnel-like section, and another on the opposite side through which the vehicles were manually reintroduced after exiting. The arena was made of wooden walls lined with a PVC sheet and magnetically attached to a metallic base plate. A glass sheet was placed 19 mm above the surface to prevent the vehicles from overturning. The exit opening was $L=30\ \mathrm{mm}$ wide, corresponding to the width of two VDVs. To measure the pressure exerted by the VDVs on the exit opening, we used two independent wooden sections [highlighted in yellow in Fig.~\ref{fig:experimental}(a)] that were connected to the main frame through two load cells (highlighted in blue). The end sections were 15 mm thick, ensuring that they did not touch either the base or the glass cover, so the forces exerted by the VDVs on them were transmitted to the load cells. The load cells had a measurement range of $\pm1$ kg. Both were connected to a 24-bit analog-to-digital converter, which allowed sampling at a rate of 44 S/s. To track the trajectories of the vehicles and determine the time at which each VDV exited the arena, we used a top-view digital video camera operating at 24 FPS. Vehicle identification was achieved using ArUco markers \cite{garrido2016generation,romero2018speeded}: each VDV was equipped with a unique tag, allowing for unambiguous particle tracking \cite{dataset}.

Each experiment lasted more than one hour and was stopped once the VDVs exiting the arena began to show signs of battery depletion. In each run, we varied the number of VDVs in the arena: $N_\mathrm{tot}=30, 40, 45,\ \mathrm{and}\ 50$. To avoid battery drain and ensure the continuity of the experiments, clogging events lasting more than two minutes were manually resolved by removing one of the VDVs involved in the clog.

\section{Results}

As a first step in our analysis, we examined how the total number of agents influences the outflow dynamics. To this end, we tracked the cumulative number of agents, $N$, exiting the arena over time for each experimental configuration, as shown in Fig.~\ref{fig:experimental}(b). As previously reported in \cite{patterson2017clogging}, increasing the number of agents in the arena leads to a reduced outflow rate. This behavior has been attributed to longer-lasting clogs, likely caused by the increased pressure within the system.

\begin{figure*}
\includegraphics{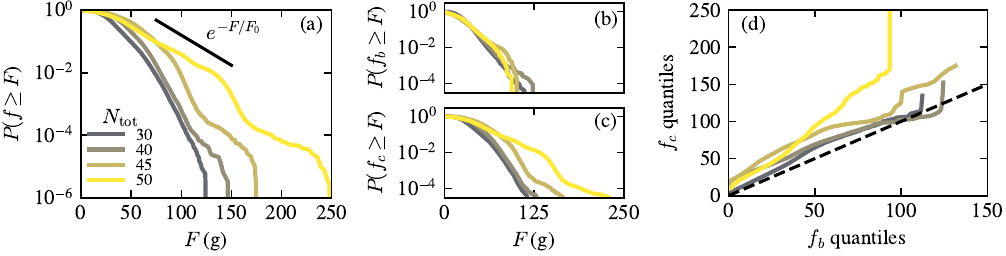}
\caption{\label{fig:forces} (a) Complementary CDF of all forces recorded with the two sensors for experiments with different number of $N_\mathrm{tot}$ as indicated in the legend. The black solid line illustrates an exponential trend. (b) and (c) show the complementary CDFs of the forces measured during bursts and clogs, respectively.
(d) Quantile-Quantile plot comparing the force distributions during bursts ($f_b$) and clogs ($f_c$). The color code in all panels is the same as in (a).}
\end{figure*}

Effectively, the analysis of the distributions of time intervals between consecutive exits of VDVs ($\Delta t$) corroborates that clog duration remarkably enlarges with the crowd size. In particular, in Fig.~\ref{fig:experimental}(c) we show the survival functions (or complementary cumulative distribution functions, CCDFs), of $\Delta t$;  i.e., the probability that an interval $\Delta t$ exceeds a given duration $\tau$, $P(\Delta t \geq \tau)$. As the number of VDVs in the system increases, the distribution tails become heavier, consistent with power-law behavior, $P(\Delta t \geq \tau) \propto \tau^{-\gamma + 1}$, which corresponds to a probability density function of the form $\mathrm{PDF}(\tau) \propto \tau^{-\gamma}$. To estimate the exponent $\gamma$, we used the method introduced by Clauset et al. \cite{clauset2009power,alstott2014powerlaw}, which yields both the power-law exponent and the minimum time $\tau_{\text{min}}$ from which the fit is valid. In our experiments, $\tau_{\text{min}}$ ranged from $3$ s to $11$ s, while the power-law exponent $\gamma$ took the values $2.2 \pm 0.2$, $2.1 \pm 0.2$, $1.7 \pm 0.2$, and $1.5 \pm 0.2$ for $N_\mathrm{tot} = 30, 40, 45,\ \mathrm{and}\ 50$, respectively. These results are consistent with those reported in Ref. \cite{patterson2017clogging}, where it was shown that increasing the number of VDVs leads to smaller values of $\gamma$, reaching values below $2$, a critical point at which the average clogging time diverges. In the particular conditions of our experiment (outlet size, VDVs level of excitation, etc.) the transition from unclogged ($\gamma > 2$) to clogged ($\gamma \leq 2$) states seems to occur in the range of $N_\mathrm{tot} = 40$ to $45$. Moreover, the value of $\tau_{\text{min}}$ estimated by the fittings can be used as a threshold to separate bursting from arrested flows. In this work, although the values of $\tau_{\text{min}}$ estimated from the power-law fits vary slightly across experimental conditions, we adopt $\tau_\mathrm{th} = 5$ s as an intermediate threshold to consistently distinguish between the bursting and arrested regimes in all experiments. The selected threshold is indicated by a vertical dashed line in Fig.~\ref{fig:experimental}(c).

Once we have confirmed that our setup reproduces the principal features of the intermittent flow observed previously in this system, we turn on the analysis of pressure measurements. Figures~\ref{fig:experimental}(d)-(e) represent typical temporal evolutions of the force signals recorded by both sensors and, superimposed, the exit times of the VDVs (red lines). Two main features are apparent from this plot: i) a noticeable correlation between the two force signals; and ii) in many cases the exit of a VDV correlates with a sudden release of the force which seems to be accompanied by a posterior slow build-up. In the following, we will carefully analyze this behavior as well as the dependence of the forces on the crowd size and the flowing state of the system (bursting or arrested).

First, we analyze the distributions of all the registered forces (grouping both sensors) for the different crowd sizes investigated. As before, aiming a better visualization of the distributions tails, we compute the CCDFs of these forces [Figure~\ref{fig:forces}(a)]. These evidence a behavior compatible with an exponential decay, a characteristic feature of granular materials and other discrete systems \cite{expforce1,expforce2}.  Also, the plots evidence that the scenarios with larger number of particles yield higher force values. 

To investigate the origin of this dependence, we discriminate the forces developed when the system is bursting ($f_b$ when $\Delta t < \tau_\mathrm{th}$) from those appearing during clogs or arrests ($f_c$ when $\Delta t > \tau_\mathrm{th}$). Figures~\ref{fig:forces}(b) and (c) display the corresponding distributions revealing that forces developed during bursts remain similar independently on the number of agents in the arena. On the contrary, forces built up during clogs depend on the crowd size suggesting that the observed differences in the global distributions of Fig.~\ref{fig:forces}(a) primarily originate from the behavior occurring at the flow arrests. Moreover, comparing the distributions during bursts and clog development for the same number of agents, we found a feature consistent with the classification of regimes based on the exponent $\gamma$ shown in Fig.~\ref{fig:experimental}(e). For $N_\mathrm{tot} = 30$ and $40$, where the system is unclogged ($\gamma > 2$), the force distributions remain similar regardless of whether there is a clog or not at the time at which the force is measured. In contrast, for $N_\mathrm{tot} = 45$ and $50$, where $\gamma \leq 2$, the system enters a clogged regime characterized by a clear difference in the force measurement depending on whether it is taken during the burst or the arrest. This behavior is also evident in the quantile-quantile plot of $f_c$ and $f_b$ [Figure~\ref{fig:forces}(d)], where the curves for $N_\mathrm{tot} = 30$ and $40$ lie close to the identity line, indicating similar force distributions. In contrast, the curves for $N_\mathrm{tot} = 45$ and $50$ show clear deviations, reflecting the development of larger forces during arrests than during bursts.

\begin{figure}
\includegraphics{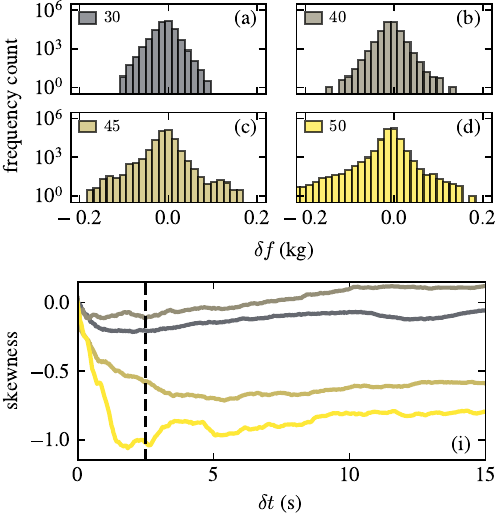}
\caption{\label{fig:skewness} (a)–(d) Histograms of the distributions of the force change $\delta f$ in a time interval $\delta t = 2.5\ \mathrm{s}$ for each of the experimental configurations. (e) Skewness of the $\delta f$ distribution as a function of $\delta t$. The vertical line indicates $\delta t = 2.5\ \mathrm{s}$, the value used in the analysis shown in panels (a)–(d). The color code in all panels is the same as in (a)–(d).}
\end{figure}

In principle, this behavior seems to be related with the dynamics already shown in Figs.~\ref{fig:experimental}(d) and (e), where sudden drops of pressure occur when an agent exits the arena, being often followed by gradual increases in pressure while the system remains clogged. Then, to gain further insight into these dynamics, we have computed the force increments $\delta f$ over a given time $\delta t$; i.e. $\delta f  = f(t + \delta t) - f(t)$. As an example, in Figs.~\ref{fig:skewness}(a)–(d) we represent the distributions of $\delta f$ using $\delta t = 2.5$~s for the four crowd sizes investigated here. The intermittent unclogged states ($N_\mathrm{tot}=30$ and $40$) yield distributions that are symmetric and narrow (with a width increasing with $N_\mathrm{tot}$). By contrast, clogged states ($N_\mathrm{tot}=45$ and $50$) exhibit moderate to strong asymmetries, with a predominance of negative $\delta f$ values. This reflects that force drops are both more frequent and more intense than force grows when the system is in a clogged state. To corroborate this result and generalize it, Fig.~\ref{fig:skewness}(e) shows the skewness of the $\delta f$ distributions as a function of $\delta t$. The unclogged scenarios ($N_\mathrm{tot}=30$ and $40$) maintain skewness values slightly below zero, indicating the existence of nearly symmetric force increment distributions regardless the time window used to compute them. This can be explained by a negligible pressure accumulation during clogs, in agreement with the results displayed in Fig.~\ref{fig:forces}(d). On the contrary, for the clogged state ($N_\mathrm{tot}=45$ and $50$), the skewness is always negative, confirming the strong asymmetry in the temporal evolution of pressure which is characterized by intense force drops and slow force growth. Interestingly, in the clogged state, the skewness becomes increasingly negative with increasing $\delta t$, saturating for $\delta t \approx 3\ \mathrm{s}$. This value suggests a characteristic timescale during which force release events take place, dominating the dynamics over another, larger timescale, at which the stress accumulation occurs.

\begin{figure*}
\includegraphics{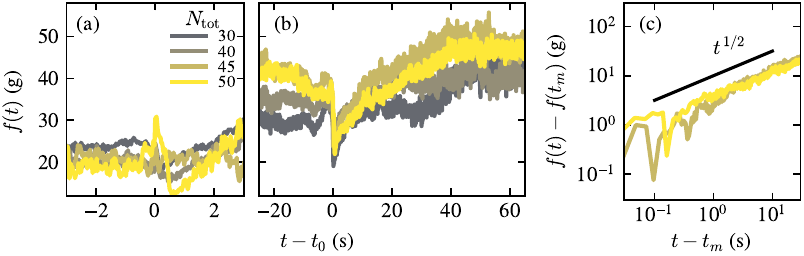}
\caption{\label{fig:failure} Average force around exit events when (a) $\Delta t\leq \tau_\mathrm{th}$ and (b) $\Delta t > \tau_\mathrm{th}$. In this case $\tau_\mathrm{th}=5$s as obtained from the survival distribution functions of Fig. \ref{fig:experimental}c. For this analysis the signal is sliced and for each event, $t_0$ is defined at the precise moment of the VDV exit time. (c) Temporal evolution of the force after the minimum force $f(t_m)$ detected during exit events in the clogging regime for $N_\mathrm{tot}=45\ \mathrm{and}\ 50$. The color code in all panels is the same as in panel (a).}
\end{figure*}

To further investigate the pressure temporal dependence on the exit of the VDVs, we sliced the pressure signal taking as $t_0$ the exit time of each agent. Figure~\ref{fig:failure}(a) shows the average force over time for events with $\Delta t < \tau_\mathrm{th}$ (with $\tau_\mathrm{th}=5$ s), so these correspond to a bursting condition as explained before. Accordingly, Fig.~\ref{fig:failure}(b) represents the average force for events with $\Delta t > \tau_\mathrm{th}$, corresponding to an arrested scenarios. In particular, for events with $\Delta t > \tau_\mathrm{th}$, the average force before an exit is obtained by averaging forces over all intervals longer than 5 s that precede an exit, whereas the average force after an exit is obtained by averaging forces over all intervals longer than 5 s that follow an exit. For events with $\Delta t < \tau_\mathrm{th}$, the same procedure applies, but considering only intervals shorter than 5 s.

In the case of bursting condition [Fig.~\ref{fig:failure}(a)], the temporal sequences are noisy, revealing a small negative peak at $t_0$, only for the case of $N_\mathrm{tot}=50$. The forces remain relatively low (around $20$ g) and do not show a clear trend before or after the exit event. In contrast, for the arrests, a clear drop of the force occurs at the exit of the agent. Following it, the force consistently grows over time, reaching values much higher than those observed in the burst cases. Importantly, a remarkable difference is observed among the unclogged ($N_\mathrm{tot}=30$ and $40$) and clogged states ($N_\mathrm{tot}=45$ and $50$), as for the latest the drop is more defined and the growth temporal dependence better resembles a square root trend.   

Aiming a better quantification of the force growing process during flow arrests in the clogged state, we compute $f(t_m)$, the force relative to the minimum value developed around $t_0$, and shift the time axis by taking $t_m$ as the reference. In other words, we align all events at the moment of minimum force, allowing a clearer comparison of the subsequent pressure increase. Then, a log-log plot of the pressure growth [Fig.~\ref{fig:failure}(c)] reveals that the force buildup clearly follows a power-law with an exponent of approximately $1/2$. This demonstrates that the pressure increases at a decelerating rate as time progresses after the minimum force event. This kind of behavior is characteristic of diffusion-like or relaxation processes, where the system gradually accumulates stress or force over time, but with diminishing increments. Such scaling can imply that the interactions or rearrangements within the system during large clog events are governed by collective dynamics that slow down the increase in pressure; a behavior that is compatible with the aging that has been proposed as the origin of the broad tails in the clogging time distributions \cite{Merrigan, GuerreroBulbul}.

\section{Conclusions}

In this work, we present an experimental study of the mechanical pressure dynamics in the flow of self-propelled particles through a narrow opening. Our results show that pressure dynamics is strongly correlated with the system’s dynamical phase; i.e. clogged or unclogged. In the clogged regime, there is a net build up of pressure after the passage of each agent, leading to a clear difference among pressures during bursts and arrests. Furthermore, the distribution of pressure increments becomes asymmetric, with a marked predominance of large negative fluctuations. In contrast, in the unclogged regime, the pressure build up becomes negligible, implying no difference among pressures during clogs and bursts, and symmetric distributions of pressure increments.

Strikingly, when the system is clogged, the pressure grows with time following a square-root dependence. This scaling is consistent with a structural aging process, previously proposed as a mechanism underlying the broad, power-law-like distribution of clogging intervals. These findings suggest that the temporal evolution of mechanical pressure can serve as a reliable indicator of the system’s state, offering an alternative to more tedious methods based on bursts and arrest intervals measurements.

\section*{ACKNOWLEDGMENTS}
This work was funded by project PICTO-2022-ITBA-00001 (Agencia Nacional de Promoci\'on Cient\'ifica y Tecnol\'ogica, Argentina and Instituto Tecnol\'ogico de Buenos Aires, Argentina) and Grant PID2023-146422NB-I00 funded by MICIU/AEI/10.13039/501100011033 and by ERDF/UE.).

\bibliography{manusBib}

\begin{thebibliography}{32}%
\makeatletter
\providecommand \@ifxundefined [1]{%
 \@ifx{#1\undefined}
}%
\providecommand \@ifnum [1]{%
 \ifnum #1\expandafter \@firstoftwo
 \else \expandafter \@secondoftwo
 \fi
}%
\providecommand \@ifx [1]{%
 \ifx #1\expandafter \@firstoftwo
 \else \expandafter \@secondoftwo
 \fi
}%
\providecommand \natexlab [1]{#1}%
\providecommand \enquote  [1]{``#1''}%
\providecommand \bibnamefont  [1]{#1}%
\providecommand \bibfnamefont [1]{#1}%
\providecommand \citenamefont [1]{#1}%
\providecommand \href@noop [0]{\@secondoftwo}%
\providecommand \href [0]{\begingroup \@sanitize@url \@href}%
\providecommand \@href[1]{\@@startlink{#1}\@@href}%
\providecommand \@@href[1]{\endgroup#1\@@endlink}%
\providecommand \@sanitize@url [0]{\catcode `\\12\catcode `\$12\catcode
  `\&12\catcode `\#12\catcode `\^12\catcode `\_12\catcode `\%12\relax}%
\providecommand \@@startlink[1]{}%
\providecommand \@@endlink[0]{}%
\providecommand \url  [0]{\begingroup\@sanitize@url \@url }%
\providecommand \@url [1]{\endgroup\@href {#1}{\urlprefix }}%
\providecommand \urlprefix  [0]{URL }%
\providecommand \Eprint [0]{\href }%
\providecommand \doibase [0]{http://dx.doi.org/}%
\providecommand \selectlanguage [0]{\@gobble}%
\providecommand \bibinfo  [0]{\@secondoftwo}%
\providecommand \bibfield  [0]{\@secondoftwo}%
\providecommand \translation [1]{[#1]}%
\providecommand \BibitemOpen [0]{}%
\providecommand \bibitemStop [0]{}%
\providecommand \bibitemNoStop [0]{.\EOS\space}%
\providecommand \EOS [0]{\spacefactor3000\relax}%
\providecommand \BibitemShut  [1]{\csname bibitem#1\endcsname}%
\let\auto@bib@innerbib\@empty
%</preamble>
\bibitem [{\citenamefont {To}\ \emph {et~al.}(2001)\citenamefont {To},
  \citenamefont {Lai},\ and\ \citenamefont {Pak}}]{SiloTO}%
  \BibitemOpen
  \bibfield  {author} {\bibinfo {author} {\bibfnamefont {K.}~\bibnamefont
  {To}}, \bibinfo {author} {\bibfnamefont {P.-Y.}\ \bibnamefont {Lai}}, \ and\
  \bibinfo {author} {\bibfnamefont {H.~K.}\ \bibnamefont {Pak}},\ }\href
  {\doibase 10.1103/PhysRevLett.86.71} {\bibfield  {journal} {\bibinfo
  {journal} {Phys. Rev. Lett.}\ }\textbf {\bibinfo {volume} {86}},\ \bibinfo
  {pages} {71} (\bibinfo {year} {2001})}\BibitemShut {NoStop}%
\bibitem [{\citenamefont {Janda}\ \emph {et~al.}(2009)\citenamefont {Janda},
  \citenamefont {Maza}, \citenamefont {Garcimart{\'\i}n}, \citenamefont {Kolb},
  \citenamefont {Lanuza},\ and\ \citenamefont {Cl{\'e}ment}}]{Janda}%
  \BibitemOpen
  \bibfield  {author} {\bibinfo {author} {\bibfnamefont {A.}~\bibnamefont
  {Janda}}, \bibinfo {author} {\bibfnamefont {D.}~\bibnamefont {Maza}},
  \bibinfo {author} {\bibfnamefont {A.}~\bibnamefont {Garcimart{\'\i}n}},
  \bibinfo {author} {\bibfnamefont {E.}~\bibnamefont {Kolb}}, \bibinfo {author}
  {\bibfnamefont {J.}~\bibnamefont {Lanuza}}, \ and\ \bibinfo {author}
  {\bibfnamefont {E.}~\bibnamefont {Cl{\'e}ment}},\ }\href@noop {} {\bibfield
  {journal} {\bibinfo  {journal} {Europhysics Letters}\ }\textbf {\bibinfo
  {volume} {87}},\ \bibinfo {pages} {24002} (\bibinfo {year}
  {2009})}\BibitemShut {NoStop}%
\bibitem [{\citenamefont {Hathcock}\ \emph {et~al.}(2025)\citenamefont
  {Hathcock}, \citenamefont {Dillavou}, \citenamefont {Hanlan}, \citenamefont
  {Durian},\ and\ \citenamefont {Tu}}]{hathcock2025stochastic}%
  \BibitemOpen
  \bibfield  {author} {\bibinfo {author} {\bibfnamefont {D.}~\bibnamefont
  {Hathcock}}, \bibinfo {author} {\bibfnamefont {S.}~\bibnamefont {Dillavou}},
  \bibinfo {author} {\bibfnamefont {J.~M.}\ \bibnamefont {Hanlan}}, \bibinfo
  {author} {\bibfnamefont {D.~J.}\ \bibnamefont {Durian}}, \ and\ \bibinfo
  {author} {\bibfnamefont {Y.}~\bibnamefont {Tu}},\ }\href@noop {} {\bibfield
  {journal} {\bibinfo  {journal} {Physical Review E}\ }\textbf {\bibinfo
  {volume} {111}},\ \bibinfo {pages} {L023404} (\bibinfo {year}
  {2025})}\BibitemShut {NoStop}%
\bibitem [{\citenamefont {Garcimart{\'\i}n}\ \emph {et~al.}(2015)\citenamefont
  {Garcimart{\'\i}n}, \citenamefont {Pastor}, \citenamefont {Ferrer},
  \citenamefont {Ramos}, \citenamefont {Mart{\'\i}n-G{\'o}mez},\ and\
  \citenamefont {Zuriguel}}]{sheep}%
  \BibitemOpen
  \bibfield  {author} {\bibinfo {author} {\bibfnamefont {A.}~\bibnamefont
  {Garcimart{\'\i}n}}, \bibinfo {author} {\bibfnamefont {J.}~\bibnamefont
  {Pastor}}, \bibinfo {author} {\bibfnamefont {L.}~\bibnamefont {Ferrer}},
  \bibinfo {author} {\bibfnamefont {J.}~\bibnamefont {Ramos}}, \bibinfo
  {author} {\bibfnamefont {C.}~\bibnamefont {Mart{\'\i}n-G{\'o}mez}}, \ and\
  \bibinfo {author} {\bibfnamefont {I.}~\bibnamefont {Zuriguel}},\ }\href@noop
  {} {\bibfield  {journal} {\bibinfo  {journal} {Physical Review E}\ }\textbf
  {\bibinfo {volume} {91}},\ \bibinfo {pages} {022808} (\bibinfo {year}
  {2015})}\BibitemShut {NoStop}%
\bibitem [{\citenamefont {Zuriguel}\ \emph {et~al.}(2014)\citenamefont
  {Zuriguel}, \citenamefont {Parisi}, \citenamefont {Hidalgo}, \citenamefont
  {Lozano}, \citenamefont {Janda}, \citenamefont {Gago}, \citenamefont
  {Peralta}, \citenamefont {Ferrer}, \citenamefont {Pugnaloni}, \citenamefont
  {Cl{\'e}ment} \emph {et~al.}}]{zuriguel2014clogging}%
  \BibitemOpen
  \bibfield  {author} {\bibinfo {author} {\bibfnamefont {I.}~\bibnamefont
  {Zuriguel}}, \bibinfo {author} {\bibfnamefont {D.~R.}\ \bibnamefont
  {Parisi}}, \bibinfo {author} {\bibfnamefont {R.~C.}\ \bibnamefont {Hidalgo}},
  \bibinfo {author} {\bibfnamefont {C.}~\bibnamefont {Lozano}}, \bibinfo
  {author} {\bibfnamefont {A.}~\bibnamefont {Janda}}, \bibinfo {author}
  {\bibfnamefont {P.~A.}\ \bibnamefont {Gago}}, \bibinfo {author}
  {\bibfnamefont {J.~P.}\ \bibnamefont {Peralta}}, \bibinfo {author}
  {\bibfnamefont {L.~M.}\ \bibnamefont {Ferrer}}, \bibinfo {author}
  {\bibfnamefont {L.~A.}\ \bibnamefont {Pugnaloni}}, \bibinfo {author}
  {\bibfnamefont {E.}~\bibnamefont {Cl{\'e}ment}},  \emph {et~al.},\
  }\href@noop {} {\bibfield  {journal} {\bibinfo  {journal} {Scientific
  reports}\ }\textbf {\bibinfo {volume} {4}},\ \bibinfo {pages} {7324}
  (\bibinfo {year} {2014})}\BibitemShut {NoStop}%
\bibitem [{\citenamefont {Zuriguel}\ \emph {et~al.}(2005)\citenamefont
  {Zuriguel}, \citenamefont {Garcimart{\'\i}n}, \citenamefont {Maza},
  \citenamefont {Pugnaloni},\ and\ \citenamefont
  {Pastor}}]{zuriguel2005exponential}%
  \BibitemOpen
  \bibfield  {author} {\bibinfo {author} {\bibfnamefont {I.}~\bibnamefont
  {Zuriguel}}, \bibinfo {author} {\bibfnamefont {A.}~\bibnamefont
  {Garcimart{\'\i}n}}, \bibinfo {author} {\bibfnamefont {D.}~\bibnamefont
  {Maza}}, \bibinfo {author} {\bibfnamefont {L.~A.}\ \bibnamefont {Pugnaloni}},
  \ and\ \bibinfo {author} {\bibfnamefont {J.}~\bibnamefont {Pastor}},\
  }\href@noop {} {\bibfield  {journal} {\bibinfo  {journal} {Physical Review
  E—Statistical, Nonlinear, and Soft Matter Physics}\ }\textbf {\bibinfo
  {volume} {71}},\ \bibinfo {pages} {051303} (\bibinfo {year}
  {2005})}\BibitemShut {NoStop}%
\bibitem [{\citenamefont {Helbing}\ \emph {et~al.}(2006)\citenamefont
  {Helbing}, \citenamefont {Johansson}, \citenamefont {Mathiesen},
  \citenamefont {Jensen},\ and\ \citenamefont {Hansen}}]{helbing2006exp}%
  \BibitemOpen
  \bibfield  {author} {\bibinfo {author} {\bibfnamefont {D.}~\bibnamefont
  {Helbing}}, \bibinfo {author} {\bibfnamefont {A.}~\bibnamefont {Johansson}},
  \bibinfo {author} {\bibfnamefont {J.}~\bibnamefont {Mathiesen}}, \bibinfo
  {author} {\bibfnamefont {M.~H.}\ \bibnamefont {Jensen}}, \ and\ \bibinfo
  {author} {\bibfnamefont {A.}~\bibnamefont {Hansen}},\ }\href@noop {}
  {\bibfield  {journal} {\bibinfo  {journal} {Physical review letters}\
  }\textbf {\bibinfo {volume} {97}},\ \bibinfo {pages} {168001} (\bibinfo
  {year} {2006})}\BibitemShut {NoStop}%
\bibitem [{\citenamefont {Roussel}\ \emph {et~al.}(2007)\citenamefont
  {Roussel}, \citenamefont {Nguyen},\ and\ \citenamefont {Coussot}}]{Roussel}%
  \BibitemOpen
  \bibfield  {author} {\bibinfo {author} {\bibfnamefont {N.}~\bibnamefont
  {Roussel}}, \bibinfo {author} {\bibfnamefont {T.~L.~H.}\ \bibnamefont
  {Nguyen}}, \ and\ \bibinfo {author} {\bibfnamefont {P.}~\bibnamefont
  {Coussot}},\ }\href {\doibase 10.1103/PhysRevLett.98.114502} {\bibfield
  {journal} {\bibinfo  {journal} {Phys. Rev. Lett.}\ }\textbf {\bibinfo
  {volume} {98}},\ \bibinfo {pages} {114502} (\bibinfo {year}
  {2007})}\BibitemShut {NoStop}%
\bibitem [{\citenamefont {Masuda}\ \emph {et~al.}(2014)\citenamefont {Masuda},
  \citenamefont {Nishinari},\ and\ \citenamefont
  {Schadschneider}}]{masuda2014critical}%
  \BibitemOpen
  \bibfield  {author} {\bibinfo {author} {\bibfnamefont {T.}~\bibnamefont
  {Masuda}}, \bibinfo {author} {\bibfnamefont {K.}~\bibnamefont {Nishinari}}, \
  and\ \bibinfo {author} {\bibfnamefont {A.}~\bibnamefont {Schadschneider}},\
  }\href@noop {} {\bibfield  {journal} {\bibinfo  {journal} {Physical review
  letters}\ }\textbf {\bibinfo {volume} {112}},\ \bibinfo {pages} {138701}
  (\bibinfo {year} {2014})}\BibitemShut {NoStop}%
\bibitem [{\citenamefont {Thomas}\ and\ \citenamefont {Durian}(2015)}]{Thomas}%
  \BibitemOpen
  \bibfield  {author} {\bibinfo {author} {\bibfnamefont {C.~C.}\ \bibnamefont
  {Thomas}}\ and\ \bibinfo {author} {\bibfnamefont {D.~J.}\ \bibnamefont
  {Durian}},\ }\href {\doibase 10.1103/PhysRevLett.114.178001} {\bibfield
  {journal} {\bibinfo  {journal} {Phys. Rev. Lett.}\ }\textbf {\bibinfo
  {volume} {114}},\ \bibinfo {pages} {178001} (\bibinfo {year}
  {2015})}\BibitemShut {NoStop}%
\bibitem [{\citenamefont {Nicolas}\ \emph {et~al.}(2018)\citenamefont
  {Nicolas}, \citenamefont {Garcimart\'{\i}n},\ and\ \citenamefont
  {Zuriguel}}]{Nicolas}%
  \BibitemOpen
  \bibfield  {author} {\bibinfo {author} {\bibfnamefont {A.}~\bibnamefont
  {Nicolas}}, \bibinfo {author} {\bibfnamefont {A.}~\bibnamefont
  {Garcimart\'{\i}n}}, \ and\ \bibinfo {author} {\bibfnamefont
  {I.}~\bibnamefont {Zuriguel}},\ }\href {\doibase
  10.1103/PhysRevLett.120.198002} {\bibfield  {journal} {\bibinfo  {journal}
  {Phys. Rev. Lett.}\ }\textbf {\bibinfo {volume} {120}},\ \bibinfo {pages}
  {198002} (\bibinfo {year} {2018})}\BibitemShut {NoStop}%
\bibitem [{\citenamefont {Merrigan}\ \emph {et~al.}(2018)\citenamefont
  {Merrigan}, \citenamefont {Birwa}, \citenamefont {Tewari},\ and\
  \citenamefont {Chakraborty}}]{Merrigan}%
  \BibitemOpen
  \bibfield  {author} {\bibinfo {author} {\bibfnamefont {C.}~\bibnamefont
  {Merrigan}}, \bibinfo {author} {\bibfnamefont {S.~K.}\ \bibnamefont {Birwa}},
  \bibinfo {author} {\bibfnamefont {S.}~\bibnamefont {Tewari}}, \ and\ \bibinfo
  {author} {\bibfnamefont {B.}~\bibnamefont {Chakraborty}},\ }\href {\doibase
  10.1103/PhysRevE.97.040901} {\bibfield  {journal} {\bibinfo  {journal} {Phys.
  Rev. E}\ }\textbf {\bibinfo {volume} {97}},\ \bibinfo {pages} {040901}
  (\bibinfo {year} {2018})}\BibitemShut {NoStop}%
\bibitem [{\citenamefont {Guerrero}\ \emph {et~al.}(2019)\citenamefont
  {Guerrero}, \citenamefont {Chakraborty}, \citenamefont {Zuriguel},\ and\
  \citenamefont {Garcimart\'{\i}n}}]{GuerreroBulbul}%
  \BibitemOpen
  \bibfield  {author} {\bibinfo {author} {\bibfnamefont {B.~V.}\ \bibnamefont
  {Guerrero}}, \bibinfo {author} {\bibfnamefont {B.}~\bibnamefont
  {Chakraborty}}, \bibinfo {author} {\bibfnamefont {I.}~\bibnamefont
  {Zuriguel}}, \ and\ \bibinfo {author} {\bibfnamefont {A.}~\bibnamefont
  {Garcimart\'{\i}n}},\ }\href {\doibase 10.1103/PhysRevE.100.032901}
  {\bibfield  {journal} {\bibinfo  {journal} {Phys. Rev. E}\ }\textbf {\bibinfo
  {volume} {100}},\ \bibinfo {pages} {032901} (\bibinfo {year}
  {2019})}\BibitemShut {NoStop}%
\bibitem [{\citenamefont {Larrieu}\ \emph {et~al.}(2023)\citenamefont
  {Larrieu}, \citenamefont {Moreau}, \citenamefont {Graff}, \citenamefont
  {Peyla},\ and\ \citenamefont {Dupont}}]{fish}%
  \BibitemOpen
  \bibfield  {author} {\bibinfo {author} {\bibfnamefont {R.}~\bibnamefont
  {Larrieu}}, \bibinfo {author} {\bibfnamefont {P.}~\bibnamefont {Moreau}},
  \bibinfo {author} {\bibfnamefont {C.}~\bibnamefont {Graff}}, \bibinfo
  {author} {\bibfnamefont {P.}~\bibnamefont {Peyla}}, \ and\ \bibinfo {author}
  {\bibfnamefont {A.}~\bibnamefont {Dupont}},\ }\href@noop {} {\bibfield
  {journal} {\bibinfo  {journal} {Scientific reports}\ }\textbf {\bibinfo
  {volume} {13}},\ \bibinfo {pages} {10414} (\bibinfo {year}
  {2023})}\BibitemShut {NoStop}%
\bibitem [{\citenamefont {Echeverr{\'\i}a-Huarte}\ \emph
  {et~al.}(2022)\citenamefont {Echeverr{\'\i}a-Huarte}, \citenamefont {Shi},
  \citenamefont {Garcimart{\'\i}n},\ and\ \citenamefont
  {Zuriguel}}]{echeverria2022pedsocial}%
  \BibitemOpen
  \bibfield  {author} {\bibinfo {author} {\bibfnamefont {I.}~\bibnamefont
  {Echeverr{\'\i}a-Huarte}}, \bibinfo {author} {\bibfnamefont {Z.}~\bibnamefont
  {Shi}}, \bibinfo {author} {\bibfnamefont {A.}~\bibnamefont
  {Garcimart{\'\i}n}}, \ and\ \bibinfo {author} {\bibfnamefont
  {I.}~\bibnamefont {Zuriguel}},\ }\href@noop {} {\bibfield  {journal}
  {\bibinfo  {journal} {Physical Review E}\ }\textbf {\bibinfo {volume}
  {106}},\ \bibinfo {pages} {044302} (\bibinfo {year} {2022})}\BibitemShut
  {NoStop}%
\bibitem [{\citenamefont {Pastor}\ \emph {et~al.}(2015)\citenamefont {Pastor},
  \citenamefont {Garcimart{\'\i}n}, \citenamefont {Gago}, \citenamefont
  {Peralta}, \citenamefont {Mart{\'\i}n-G{\'o}mez}, \citenamefont {Ferrer},
  \citenamefont {Maza}, \citenamefont {Parisi}, \citenamefont {Pugnaloni},\
  and\ \citenamefont {Zuriguel}}]{fasterisslower}%
  \BibitemOpen
  \bibfield  {author} {\bibinfo {author} {\bibfnamefont {J.~M.}\ \bibnamefont
  {Pastor}}, \bibinfo {author} {\bibfnamefont {A.}~\bibnamefont
  {Garcimart{\'\i}n}}, \bibinfo {author} {\bibfnamefont {P.~A.}\ \bibnamefont
  {Gago}}, \bibinfo {author} {\bibfnamefont {J.~P.}\ \bibnamefont {Peralta}},
  \bibinfo {author} {\bibfnamefont {C.}~\bibnamefont {Mart{\'\i}n-G{\'o}mez}},
  \bibinfo {author} {\bibfnamefont {L.~M.}\ \bibnamefont {Ferrer}}, \bibinfo
  {author} {\bibfnamefont {D.}~\bibnamefont {Maza}}, \bibinfo {author}
  {\bibfnamefont {D.~R.}\ \bibnamefont {Parisi}}, \bibinfo {author}
  {\bibfnamefont {L.~A.}\ \bibnamefont {Pugnaloni}}, \ and\ \bibinfo {author}
  {\bibfnamefont {I.}~\bibnamefont {Zuriguel}},\ }\href@noop {} {\bibfield
  {journal} {\bibinfo  {journal} {Physical Review E}\ }\textbf {\bibinfo
  {volume} {92}},\ \bibinfo {pages} {062817} (\bibinfo {year}
  {2015})}\BibitemShut {NoStop}%
\bibitem [{\citenamefont {Souzy}\ and\ \citenamefont
  {Marin}(2022)}]{Souzy_Marin_2022}%
  \BibitemOpen
  \bibfield  {author} {\bibinfo {author} {\bibfnamefont {M.}~\bibnamefont
  {Souzy}}\ and\ \bibinfo {author} {\bibfnamefont {A.}~\bibnamefont {Marin}},\
  }\href {\doibase 10.1017/jfm.2022.981} {\bibfield  {journal} {\bibinfo
  {journal} {Journal of Fluid Mechanics}\ }\textbf {\bibinfo {volume} {953}},\
  \bibinfo {pages} {A40} (\bibinfo {year} {2022})}\BibitemShut {NoStop}%
\bibitem [{\citenamefont {Patterson}\ \emph {et~al.}(2017)\citenamefont
  {Patterson}, \citenamefont {Fierens}, \citenamefont {Sangiuliano~Jimka},
  \citenamefont {K\"onig}, \citenamefont {Garcimart\'{\i}n}, \citenamefont
  {Zuriguel}, \citenamefont {Pugnaloni},\ and\ \citenamefont
  {Parisi}}]{patterson2017clogging}%
  \BibitemOpen
  \bibfield  {author} {\bibinfo {author} {\bibfnamefont {G.~A.}\ \bibnamefont
  {Patterson}}, \bibinfo {author} {\bibfnamefont {P.~I.}\ \bibnamefont
  {Fierens}}, \bibinfo {author} {\bibfnamefont {F.}~\bibnamefont
  {Sangiuliano~Jimka}}, \bibinfo {author} {\bibfnamefont {P.~G.}\ \bibnamefont
  {K\"onig}}, \bibinfo {author} {\bibfnamefont {A.}~\bibnamefont
  {Garcimart\'{\i}n}}, \bibinfo {author} {\bibfnamefont {I.}~\bibnamefont
  {Zuriguel}}, \bibinfo {author} {\bibfnamefont {L.~A.}\ \bibnamefont
  {Pugnaloni}}, \ and\ \bibinfo {author} {\bibfnamefont {D.~R.}\ \bibnamefont
  {Parisi}},\ }\href {\doibase 10.1103/PhysRevLett.119.248301} {\bibfield
  {journal} {\bibinfo  {journal} {Phys. Rev. Lett.}\ }\textbf {\bibinfo
  {volume} {119}},\ \bibinfo {pages} {248301} (\bibinfo {year}
  {2017})}\BibitemShut {NoStop}%
\bibitem [{\citenamefont {Marin}\ and\ \citenamefont
  {Souzy}(2024)}]{marin2024review}%
  \BibitemOpen
  \bibfield  {author} {\bibinfo {author} {\bibfnamefont {A.}~\bibnamefont
  {Marin}}\ and\ \bibinfo {author} {\bibfnamefont {M.}~\bibnamefont {Souzy}},\
  }\href@noop {} {\bibfield  {journal} {\bibinfo  {journal} {Annual Review of
  Fluid Mechanics}\ }\textbf {\bibinfo {volume} {57}} (\bibinfo {year}
  {2024})}\BibitemShut {NoStop}%
\bibitem [{\citenamefont {Zuriguel}\ \emph {et~al.}(2020)\citenamefont
  {Zuriguel}, \citenamefont {Echeverr{\'\i}a}, \citenamefont {Maza},
  \citenamefont {Hidalgo}, \citenamefont {Mart{\'\i}n-G{\'o}mez},\ and\
  \citenamefont {Garcimart{\'\i}n}}]{pressurepeds}%
  \BibitemOpen
  \bibfield  {author} {\bibinfo {author} {\bibfnamefont {I.}~\bibnamefont
  {Zuriguel}}, \bibinfo {author} {\bibfnamefont {I.}~\bibnamefont
  {Echeverr{\'\i}a}}, \bibinfo {author} {\bibfnamefont {D.}~\bibnamefont
  {Maza}}, \bibinfo {author} {\bibfnamefont {R.~C.}\ \bibnamefont {Hidalgo}},
  \bibinfo {author} {\bibfnamefont {C.}~\bibnamefont {Mart{\'\i}n-G{\'o}mez}},
  \ and\ \bibinfo {author} {\bibfnamefont {A.}~\bibnamefont
  {Garcimart{\'\i}n}},\ }\href@noop {} {\bibfield  {journal} {\bibinfo
  {journal} {Safety science}\ }\textbf {\bibinfo {volume} {121}},\ \bibinfo
  {pages} {394} (\bibinfo {year} {2020})}\BibitemShut {NoStop}%
\bibitem [{\citenamefont {Longhi}\ \emph {et~al.}(2002)\citenamefont {Longhi},
  \citenamefont {Easwar},\ and\ \citenamefont {Menon}}]{expforce2}%
  \BibitemOpen
  \bibfield  {author} {\bibinfo {author} {\bibfnamefont {E.}~\bibnamefont
  {Longhi}}, \bibinfo {author} {\bibfnamefont {N.}~\bibnamefont {Easwar}}, \
  and\ \bibinfo {author} {\bibfnamefont {N.}~\bibnamefont {Menon}},\
  }\href@noop {} {\bibfield  {journal} {\bibinfo  {journal} {Physical review
  letters}\ }\textbf {\bibinfo {volume} {89}},\ \bibinfo {pages} {045501}
  (\bibinfo {year} {2002})}\BibitemShut {NoStop}%
\bibitem [{\citenamefont {Barois}\ \emph {et~al.}(2019)\citenamefont {Barois},
  \citenamefont {Boudet}, \citenamefont {Lanchon}, \citenamefont {Lintuvuori},\
  and\ \citenamefont {Kellay}}]{selfbottleneck}%
  \BibitemOpen
  \bibfield  {author} {\bibinfo {author} {\bibfnamefont {T.}~\bibnamefont
  {Barois}}, \bibinfo {author} {\bibfnamefont {J.-F. m.~c.}\ \bibnamefont
  {Boudet}}, \bibinfo {author} {\bibfnamefont {N.}~\bibnamefont {Lanchon}},
  \bibinfo {author} {\bibfnamefont {J.~S.}\ \bibnamefont {Lintuvuori}}, \ and\
  \bibinfo {author} {\bibfnamefont {H.}~\bibnamefont {Kellay}},\ }\href
  {\doibase 10.1103/PhysRevE.99.052605} {\bibfield  {journal} {\bibinfo
  {journal} {Phys. Rev. E}\ }\textbf {\bibinfo {volume} {99}},\ \bibinfo
  {pages} {052605} (\bibinfo {year} {2019})}\BibitemShut {NoStop}%
\bibitem [{\citenamefont {Deblais}\ \emph {et~al.}(2018)\citenamefont
  {Deblais}, \citenamefont {Barois}, \citenamefont {Guerin}, \citenamefont
  {Delville}, \citenamefont {Vaudaine}, \citenamefont {Lintuvuori},
  \citenamefont {Boudet}, \citenamefont {Baret},\ and\ \citenamefont
  {Kellay}}]{self3}%
  \BibitemOpen
  \bibfield  {author} {\bibinfo {author} {\bibfnamefont {A.}~\bibnamefont
  {Deblais}}, \bibinfo {author} {\bibfnamefont {T.}~\bibnamefont {Barois}},
  \bibinfo {author} {\bibfnamefont {T.}~\bibnamefont {Guerin}}, \bibinfo
  {author} {\bibfnamefont {P.-H.}\ \bibnamefont {Delville}}, \bibinfo {author}
  {\bibfnamefont {R.}~\bibnamefont {Vaudaine}}, \bibinfo {author}
  {\bibfnamefont {J.~S.}\ \bibnamefont {Lintuvuori}}, \bibinfo {author}
  {\bibfnamefont {J.-F.}\ \bibnamefont {Boudet}}, \bibinfo {author}
  {\bibfnamefont {J.-C.}\ \bibnamefont {Baret}}, \ and\ \bibinfo {author}
  {\bibfnamefont {H.}~\bibnamefont {Kellay}},\ }\href@noop {} {\bibfield
  {journal} {\bibinfo  {journal} {Physical review letters}\ }\textbf {\bibinfo
  {volume} {120}},\ \bibinfo {pages} {188002} (\bibinfo {year}
  {2018})}\BibitemShut {NoStop}%
\bibitem [{\citenamefont {Dauchot}\ and\ \citenamefont
  {D\'emery}(2019)}]{self1}%
  \BibitemOpen
  \bibfield  {author} {\bibinfo {author} {\bibfnamefont {O.}~\bibnamefont
  {Dauchot}}\ and\ \bibinfo {author} {\bibfnamefont {V.}~\bibnamefont
  {D\'emery}},\ }\href {\doibase 10.1103/PhysRevLett.122.068002} {\bibfield
  {journal} {\bibinfo  {journal} {Phys. Rev. Lett.}\ }\textbf {\bibinfo
  {volume} {122}},\ \bibinfo {pages} {068002} (\bibinfo {year}
  {2019})}\BibitemShut {NoStop}%
\bibitem [{\citenamefont {Barois}\ \emph {et~al.}(2020)\citenamefont {Barois},
  \citenamefont {Boudet}, \citenamefont {Lintuvuori},\ and\ \citenamefont
  {Kellay}}]{self4}%
  \BibitemOpen
  \bibfield  {author} {\bibinfo {author} {\bibfnamefont {T.}~\bibnamefont
  {Barois}}, \bibinfo {author} {\bibfnamefont {J.-F. m.~c.}\ \bibnamefont
  {Boudet}}, \bibinfo {author} {\bibfnamefont {J.~S.}\ \bibnamefont
  {Lintuvuori}}, \ and\ \bibinfo {author} {\bibfnamefont {H.}~\bibnamefont
  {Kellay}},\ }\href {\doibase 10.1103/PhysRevLett.125.238003} {\bibfield
  {journal} {\bibinfo  {journal} {Phys. Rev. Lett.}\ }\textbf {\bibinfo
  {volume} {125}},\ \bibinfo {pages} {238003} (\bibinfo {year}
  {2020})}\BibitemShut {NoStop}%
\bibitem [{\citenamefont {Baconnier}\ \emph {et~al.}(2025)\citenamefont
  {Baconnier}, \citenamefont {Dauchot}, \citenamefont {D\'emery}, \citenamefont
  {D\"uring}, \citenamefont {Henkes}, \citenamefont {Huepe},\ and\
  \citenamefont {Shee}}]{self2}%
  \BibitemOpen
  \bibfield  {author} {\bibinfo {author} {\bibfnamefont {P.}~\bibnamefont
  {Baconnier}}, \bibinfo {author} {\bibfnamefont {O.}~\bibnamefont {Dauchot}},
  \bibinfo {author} {\bibfnamefont {V.}~\bibnamefont {D\'emery}}, \bibinfo
  {author} {\bibfnamefont {G.}~\bibnamefont {D\"uring}}, \bibinfo {author}
  {\bibfnamefont {S.}~\bibnamefont {Henkes}}, \bibinfo {author} {\bibfnamefont
  {C.}~\bibnamefont {Huepe}}, \ and\ \bibinfo {author} {\bibfnamefont
  {A.}~\bibnamefont {Shee}},\ }\href {\doibase 10.1103/RevModPhys.97.015007}
  {\bibfield  {journal} {\bibinfo  {journal} {Rev. Mod. Phys.}\ }\textbf
  {\bibinfo {volume} {97}},\ \bibinfo {pages} {015007} (\bibinfo {year}
  {2025})}\BibitemShut {NoStop}%
\bibitem [{\citenamefont {Garrido-Jurado}\ \emph {et~al.}(2016)\citenamefont
  {Garrido-Jurado}, \citenamefont {Muñoz-Salinas}, \citenamefont
  {Madrid-Cuevas},\ and\ \citenamefont
  {Medina-Carnicer}}]{garrido2016generation}%
  \BibitemOpen
  \bibfield  {author} {\bibinfo {author} {\bibfnamefont {S.}~\bibnamefont
  {Garrido-Jurado}}, \bibinfo {author} {\bibfnamefont {R.}~\bibnamefont
  {Muñoz-Salinas}}, \bibinfo {author} {\bibfnamefont {F.}~\bibnamefont
  {Madrid-Cuevas}}, \ and\ \bibinfo {author} {\bibfnamefont {R.}~\bibnamefont
  {Medina-Carnicer}},\ }\href {\doibase
  https://doi.org/10.1016/j.patcog.2015.09.023} {\bibfield  {journal} {\bibinfo
   {journal} {Pattern Recognition}\ }\textbf {\bibinfo {volume} {51}},\
  \bibinfo {pages} {481} (\bibinfo {year} {2016})}\BibitemShut {NoStop}%
\bibitem [{\citenamefont {Romero-Ramirez}\ \emph {et~al.}(2018)\citenamefont
  {Romero-Ramirez}, \citenamefont {Muñoz-Salinas},\ and\ \citenamefont
  {Medina-Carnicer}}]{romero2018speeded}%
  \BibitemOpen
  \bibfield  {author} {\bibinfo {author} {\bibfnamefont {F.~J.}\ \bibnamefont
  {Romero-Ramirez}}, \bibinfo {author} {\bibfnamefont {R.}~\bibnamefont
  {Muñoz-Salinas}}, \ and\ \bibinfo {author} {\bibfnamefont {R.}~\bibnamefont
  {Medina-Carnicer}},\ }\href {\doibase
  https://doi.org/10.1016/j.imavis.2018.05.004} {\bibfield  {journal} {\bibinfo
   {journal} {Image and Vision Computing}\ }\textbf {\bibinfo {volume} {76}},\
  \bibinfo {pages} {38} (\bibinfo {year} {2018})}\BibitemShut {NoStop}%
\bibitem [{\citenamefont {Ramdan~Ferressini}\ \emph {et~al.}(2025)\citenamefont
  {Ramdan~Ferressini}, \citenamefont {Colantuono}, \citenamefont {Zuriguel},
  \citenamefont {Parisi},\ and\ \citenamefont {Patterson}}]{dataset}%
  \BibitemOpen
  \bibfield  {author} {\bibinfo {author} {\bibfnamefont {M.}~\bibnamefont
  {Ramdan~Ferressini}}, \bibinfo {author} {\bibfnamefont {N.}~\bibnamefont
  {Colantuono}}, \bibinfo {author} {\bibfnamefont {I.}~\bibnamefont
  {Zuriguel}}, \bibinfo {author} {\bibfnamefont {D.~R.}\ \bibnamefont
  {Parisi}}, \ and\ \bibinfo {author} {\bibfnamefont {G.~A.}\ \bibnamefont
  {Patterson}},\ }\href {\doibase 10.6084/m9.figshare.29395898} {\enquote
  {\bibinfo {title} {Pressure dynamics in the bottleneck flow of self-propelled
  particles - dataset},}\ } (\bibinfo {year} {2025})\BibitemShut {NoStop}%
\bibitem [{\citenamefont {Clauset}\ \emph {et~al.}(2009)\citenamefont
  {Clauset}, \citenamefont {Shalizi},\ and\ \citenamefont
  {Newman}}]{clauset2009power}%
  \BibitemOpen
  \bibfield  {author} {\bibinfo {author} {\bibfnamefont {A.}~\bibnamefont
  {Clauset}}, \bibinfo {author} {\bibfnamefont {C.~R.}\ \bibnamefont
  {Shalizi}}, \ and\ \bibinfo {author} {\bibfnamefont {M.~E.~J.}\ \bibnamefont
  {Newman}},\ }\href {\doibase 10.1137/070710111} {\bibfield  {journal}
  {\bibinfo  {journal} {SIAM Review}\ }\textbf {\bibinfo {volume} {51}},\
  \bibinfo {pages} {661} (\bibinfo {year} {2009})}\BibitemShut {NoStop}%
\bibitem [{\citenamefont {Alstott}\ \emph {et~al.}(2014)\citenamefont
  {Alstott}, \citenamefont {Bullmore},\ and\ \citenamefont
  {Plenz}}]{alstott2014powerlaw}%
  \BibitemOpen
  \bibfield  {author} {\bibinfo {author} {\bibfnamefont {J.}~\bibnamefont
  {Alstott}}, \bibinfo {author} {\bibfnamefont {E.}~\bibnamefont {Bullmore}}, \
  and\ \bibinfo {author} {\bibfnamefont {D.}~\bibnamefont {Plenz}},\ }\href
  {\doibase 10.1371/journal.pone.0085777} {\bibfield  {journal} {\bibinfo
  {journal} {PLOS ONE}\ }\textbf {\bibinfo {volume} {9}},\ \bibinfo {pages} {1}
  (\bibinfo {year} {2014})}\BibitemShut {NoStop}%
\bibitem [{\citenamefont {Radjai}\ \emph {et~al.}(1996)\citenamefont {Radjai},
  \citenamefont {Jean}, \citenamefont {Moreau},\ and\ \citenamefont
  {Roux}}]{expforce1}%
  \BibitemOpen
  \bibfield  {author} {\bibinfo {author} {\bibfnamefont {F.}~\bibnamefont
  {Radjai}}, \bibinfo {author} {\bibfnamefont {M.}~\bibnamefont {Jean}},
  \bibinfo {author} {\bibfnamefont {J.-J.}\ \bibnamefont {Moreau}}, \ and\
  \bibinfo {author} {\bibfnamefont {S.}~\bibnamefont {Roux}},\ }\href@noop {}
  {\bibfield  {journal} {\bibinfo  {journal} {Physical review letters}\
  }\textbf {\bibinfo {volume} {77}},\ \bibinfo {pages} {274} (\bibinfo {year}
  {1996})}\BibitemShut {NoStop}%
\end{thebibliography}%

\end{document}